\documentclass{jfm}
\usepackage{graphicx}
\usepackage{marginnote}
\usepackage{color}
\usepackage{amsmath}
\usepackage{tikz}
\usepackage{pgfcore}
\usepackage{caption}
\usepackage{subcaption}
\usepackage{multirow}
\usepackage{array}

\begin{document}

\makeatletter
\def\pgf@plot@curveto@handler@finish{%
  \ifpgf@plot@started%
    \pgfpathcurvebetweentimecontinue{0}{0.995}{\pgf@plot@curveto@first}{\pgf@plot@curveto@first@support}{\pgf@plot@curveto@second}{\pgf@plot@curveto@second}%
  \fi%
}

\newtheorem{lemma}{Lemma}
\newtheorem{corollary}{Corollary}
\newcolumntype{M}[1]{>{\centering\arraybackslash}m{#1}}

\shorttitle{Active attenuation of a trailing vortex} 
\shortauthor{A. M. Edstrand et al.} 

\title{Active attenuation of a trailing vortex inspired by a parabolized stability analysis}

\author	
 {
 Adam M. Edstrand\aff{1}, 
  Yiyang Sun\aff{1},
  Peter J. Schmid\aff{2},\\
  Kunihiko Taira\aff{1}, 
  \and 
  Louis N. Cattafesta III\aff{1}
  \corresp{\email{lcattafesta@fsu.edu}}
  }

\affiliation
{
\aff{1}
Department of Mechanical Engineering, Florida Center for Advanced Aero-Propulsion, \\
Florida State University, Tallahassee, FL 32310, USA
\aff{2}
Department of Mathematics, Imperial College London, London SW7 2AZ, UK
}

\maketitle


\begin{abstract}
Designing effective control for complex three-dimensional flow fields proves to be non-trivial.  Oftentimes, intuitive control strategies lead to suboptimal control. To navigate the control space, we utilize a linear parabolized stability analysis to guide the design of a control scheme for a trailing vortex flow field aft of a NACA0012 half-wing at an angle of attack $\alpha = 5^\circ$ and a chord-based Reynolds number $Re = 1000$. The stability results show that the unstable mode with the smallest growth rate (fifth wake mode) provides a pathway to excite a vortex instability, whereas the principal unstable mode remains in the wake of the wing. Inspired by this finding, we perform direct numerical simulations that excite each mode with body forces matching the shape function from the stability analysis. Relative to the baseline uncontrolled case, the principal wake mode reduces the vortex length, while the fifth wake mode further shortens the tip vortex.  Analogously, the streamwise circulation of the trailing vortex is found to be significantly reduced.  From these results, we conclude that a rudimentary linear stability analysis can provide key insight into the underlying physics and help engineers design more effective control.
\end{abstract}

\section{Introduction}
\label{sec:Intro}
The ability to control a flow has significant economic and safety impacts 
in transportation and military industries, as it can offer solutions to reduce drag and
increase the efficiency of virtually every vehicle designed. However, effective
flow control of three-dimensional flow fields with complex geometries presents
an immense challenge to control designers. Intuitive design can often lead
control designers astray from more effective but often obscure control
methodologies caused by the challenging nature of the disturbance development.
While using a disturbance to beneficially modify a flow field is the primary
focus of flow control, hydrodynamic stability theory is defined as the study of a
disturbance on an equilibrium state. The commonality between these two fields
helps strengthen the ability of a control designer to create effective control strategies. Here, we use the results from a parabolized stability
analysis to guide us in designing an effective control strategy for a trailing vortex flow field
showing that even a rudimentary linear stability analysis can provide key
insights into achieving enhancement in effective control.

Trailing vortices produce adverse effects in aeronautic, automotive, and maritime applications. A natural byproduct of lift, trailing vortices result from pressure differences between the upper and lower parts of a lifting surface. The resulting persistent vortex wake poses a potential hazard on a following aircraft, requiring conservative spacing limits imposed by air traffic control and causing air traffic delays \citep{Spalart:1998dl}. From an economic perspective, trailing vortices also produce induced drag, which accounts for a significant portion of drag on aircraft and automotive vehicles and a corresponding increased fuel consumption. Trailing vortices from trailing-edge flaps can also produce significant noise radiation, commonly termed ``flap-edge noise" \citep{Streett:2003vx}.  

Intuitive control approaches of a tip vortex lead control engineers to actuate in the tip region of the control surface.
Examples of this type of control use momentum injection through different configuration
slots in the wing tip region~\citep{Margaris:2006,Margaris:2010,
Edstrand:2015}. Alternative control strategies have modified the lift
distribution through rapidly actuated miniature Gurney flaps along 
the trailing edge~\citep{Matalanis:2007} and through trailing-edge blowing \citep{Taira:AIAAJ09}. Following another concept, trailing 
vortex wandering~\citep{Devenport:1996}, which has been attributed to a vortex instability~\citep{Jacquin:2001, Edstrand:2016}, could potentially be 
leveraged to accelerate the break up of these vortices. With the help of stability theory, the underlying physical mechanisms of these innovative, albeit intuitive strategies can be further understood and potentially be harnessed for effective flow manipulation. \cite{Crouch:2005ce} summarizes recent control efforts that either excite three-dimensional instabilities using some form of active control or attempt to alter the structure of the vortices or the vortex wake. Our physics-based approach based on a parabolized stability analysis falls into the first category.

This paper first summarizes the parabolized stability
analysis and corresponding direct numerical simulations over a half-span NACA0012 model. The parabolized 
stability results then provide nonintuitive physical insight that inspires a novel control approach for the attenuation of the vortex wake.
Direct numerical simulations are again used to test this design concept. We then discuss the impact and salient features of 
these results, followed by conclusions and future recommendations.


\section{Approach}
We consider the flow over a finite wing based on a NACA0012 airfoil
with a full-span aspect ratio of $2.5$ and an angle of attack of
$\alpha = 5^\circ.$ The Reynolds number, defined by the free-stream
velocity $U_\infty$ and the chord $c,$ is $Re = 1000.$ The variables
$U_\infty$ and $c$ are also used to render all variables
dimensionless. In this section, we state the governing equations
underlying our control design and outline the computational setup for
our analysis.

\subsection{Parabolized stability analysis}
\label{sec:PSEApproach}
The geometric configuration of a finite wing, coupled with our focus
on the interaction of the downstream wake and trailing vortex, allows
a linear framework for the description of the perturbation dynamics. A
further simplification is linked to the quasi-two dimensional nature
of the flow, where the base flow is assumed to vary only slowly in the
streamwise direction. The latter simplification motivates the
formulation of the perturbation dynamics by the Parabolized Stability
Equations (PSE), expressed as an initial value problem in the
streamwise direction~\citep{Herbert:1997}.

The linear assumption introduces the decomposition of the state vector
${\boldsymbol{\tilde{q}}}$ into a base  ${\boldsymbol{\tilde{Q}}}$
and a perturbation ${\boldsymbol{\hat{q}}}$ of infinitesimal
amplitude. The base component ${\boldsymbol{\tilde{Q}}}$ satisfies the
steady, nonlinear Navier-Stokes equations; the perturbation part is
governed by the linearized Navier-Stokes equations. Prompted by the
slow variation of the base flow in the streamwise direction, we assume
the perturbation state vector ${\boldsymbol{\hat{q}}}$ in the form
\begin{equation}
  {\boldsymbol{\hat{q}}}(\boldsymbol{x}, t) =
  \boldsymbol{q}(\epsilon x, y, z)
  \exp{\left( i\int_x \alpha(\xi)\,\text{d}\xi
    - i\omega t \right) }.
  \label{eq:DistDecomposition}
\end{equation}
This decomposition consists of a product of a slowly-varying shape
function ${\boldsymbol{q}}$ and a rapidly-varying phase
function. Moreover, the perturbation has been assumed periodic in
time, and a Fourier transform in this direction introduces the
real-valued frequency $\omega.$ The small parameter $\epsilon$ is
taken as $\mathcal{O}(Re^{-1}),$ and the local streamwise wavenumber
$\alpha$ is allowed to slowly vary with $x.$ Upon substitution of the
above decomposition~(\ref{eq:DistDecomposition}) into the linearized
Navier-Stokes equations, we can neglect $\mathcal{O}(\epsilon^2)$
elliptic terms in $x,$ and thus parabolize the governing equations in
the streamwise direction. Formally, we can then recast the governing
equations as an initial value problem shown as:
\begin{equation}
\begin{split}
  \mathbf{B}\frac{\partial \boldsymbol{q}}{\partial x} &= 
    \mathbf{A}\boldsymbol{q} \\
  \frac{\partial \boldsymbol{q}}{\partial x} &= 
    \mathbf{B}^{-1}\mathbf{A}\boldsymbol{q} =
    \mathbf{L}\boldsymbol{q},
\end{split}
  \label{eq:PSEMatrixForm}
\end{equation}
which constitutes an evolution equation in $x$ and can hence be
marched in the streamwise coordinate direction. The particular form of the matrix ${\mathbf{L}}$ is omitted for brevity but can be derived following~\cite{Schmid:2001} while retaining the spanwise $z$-derivatives.

The decomposition~(\ref{eq:DistDecomposition}) into shape and phase
functions is ambiguous and requires further specifications to render it
unique. Commonly, an auxiliary equation of the form
\begin{equation}
  \iint_S {\boldsymbol{q}}^H \frac{\partial {\boldsymbol{q}}}{\partial
    x}\,\text{d}S = \iint_S \frac{\partial}{\partial x}\left(
  \frac{1}{2} {\boldsymbol{q}}^H {\boldsymbol{q}}\right) \,\text{d} S
  = 0
  \label{eq:AuxiliaryEquation}
\end{equation}
is adopted, with $S$ denoting the $y$-$z$ plane. Physically,
this equation enforces that the kinetic energy of the shape function
is invariant under a downstream translation. Mathematically, it
enforces orthogonality between the shape function and its streamwise
derivative, and thus retains only the non-exponential behavior (in $x$) in
the shape function; the exponential components are relegated to the (rapidly-varying) phase function. Equations~(\ref{eq:PSEMatrixForm})
and~(\ref{eq:AuxiliaryEquation}) have to be solved simultaneously at
each position in $x,$ as the perturbation state vector is marched
downstream in the streamwise direction. A spectral collocation method is used in the cross-stream directions and backward difference in the streamwise direction to solve the initial-value problem numerically. To ensure grid independence, a grid resolution check has been performed, ensuring that $81\times81$ Chebyshev collocation points provide sufficient spatial resolution~\citep{Edstrand:JFM18}.

\subsection{Direct numerical simulations}

\begin{figure}
\begin{center}
\includegraphics[width=0.99\textwidth]{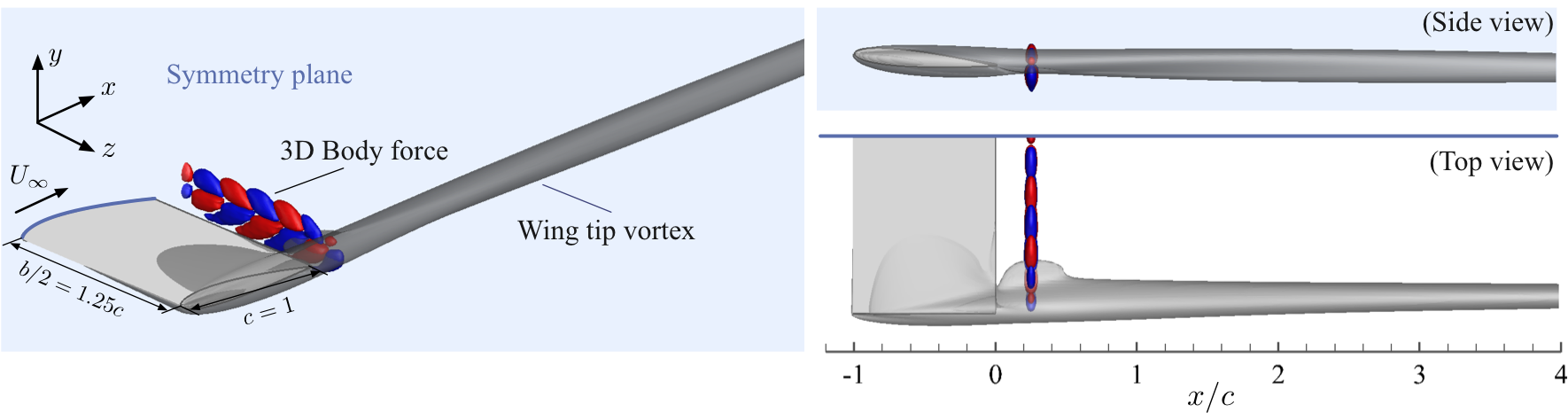}
\caption{Schematic of trailing vortex wake of a three-dimensional half-span airfoil with NACA0012 profile. Isosurfaces of streamwise vorticity $\tilde \omega_x=-0.8$ (transparent gray) and streamwise velocity perturbation (blue and red) of body force are visualized.}
\label{fig:Setup}
\end{center}
\end{figure}

Direct numerical simulations are performed for flow over a half-span
wing using the incompressible flow solver {\it Cliff} (from the {\it CharLES}
software suite \citep{Ham:CTR04,Ham:CTR06}). They complement and
support the modal analysis based on the PSE. Second-order finite-volume and
time-integration schemes are used to solve the discretized Navier--Stokes
equations. The computational setup is shown in figure~\ref{fig:Setup}. The
origin of the Cartesian coordinate system is placed at the trailing edge of the 
wing tip while $x$-, $y$-, and $z$-directions represent the streamwise, transverse,
and spanwise directions, respectively. A computational domain of
$[x/c,y/c,z/c]\in[-15,15]\times[-15,15]\times[-1.25,14.375]$ is
used. In the present numerical study, we only consider the half-span
($b/2c=1.25$) of the wing and impose symmetry boundary conditions as
illustrated in figure~\ref{fig:Setup}. At the inlet, a free stream
velocity vector of $[\tilde u,\tilde v,\tilde w] = [U_\infty,0,0]$ is prescribed and, at
the outlet, convective boundary conditions are specified. For all
other far-field boundaries, slip conditions are employed. The mesh
contains approximately 3.5 million volume cells, which is based upon
our previous work~\citep{Edstrand:JFM18} that contains a grid
resolution study to ensure adequate spatial resolution. 

With an objective to attenuate tip-vortex strength, we introduce a
modal-based body force for flow control, guided by the present parabolized
stability analysis. We use the insights from the parabolized stability analysis, such as the
spatial perturbation distribution for $\boldsymbol{q}$ and the
associated temporal frequency $\omega$ and streamwise wavenumber
$\alpha$ to introduce a three-dimensional body force
$\boldsymbol{f}_q$ expressed as
\begin{equation}
  \boldsymbol{f}_q(x,y,z,t) = A \omega \boldsymbol{q}_r(y,z) G(x)
  \sin(\omega t +\phi),
  \label{Eq:BodyForce}
\end{equation}
where $A=0.5$ is the forcing amplitude and $\phi$ is a reference
phase. The values of $\omega$ and the real component of mode
$\boldsymbol{q}_r$ are derived from the parabolized stability analysis. In
our case, the mode shape function $\boldsymbol{q}_r$ only consists of velocity
disturbances $[u,v,w]$ with unit magnitude ($||\boldsymbol
{q}_r(y,z)|| = 1$). A Gaussian distribution $G(x)$ of the form
\begin{equation}
  G(x) = \frac{1}{\sqrt{2\pi\sigma^2}}e^{-\frac{(x-0.25c)^2}{2\sigma^2}}
  \label{Eq:Gaussian}
\end{equation}
places the body force one quarter-chord aft of the trailing edge with a variance
$\sigma^2=4\times10^{-4}$ providing a sufficiently compact forcing to
render negligible the influence of the streamwise wavenumber $\alpha$
on the chosen force $\boldsymbol{f}_q$ in Eq.~(\ref{Eq:BodyForce}).

For controlled flow simulations, the mesh is further refined in two
regions to ensure accurate results: the first region is around
$x/c=0.25$ where the body force is introduced, and the second region
covers the wake zone downstream of the trailing edge. The refined mesh
contains approximately 5 million cells and has been verified (through
additional grid resolution studies) to capture all relevant features
of the unsteady controlled flows.

\section{PSE-analysis of the wake}
\label{sec:pse}

\begin{figure}
  \begin{center}
    \includegraphics[width=0.99\textwidth]{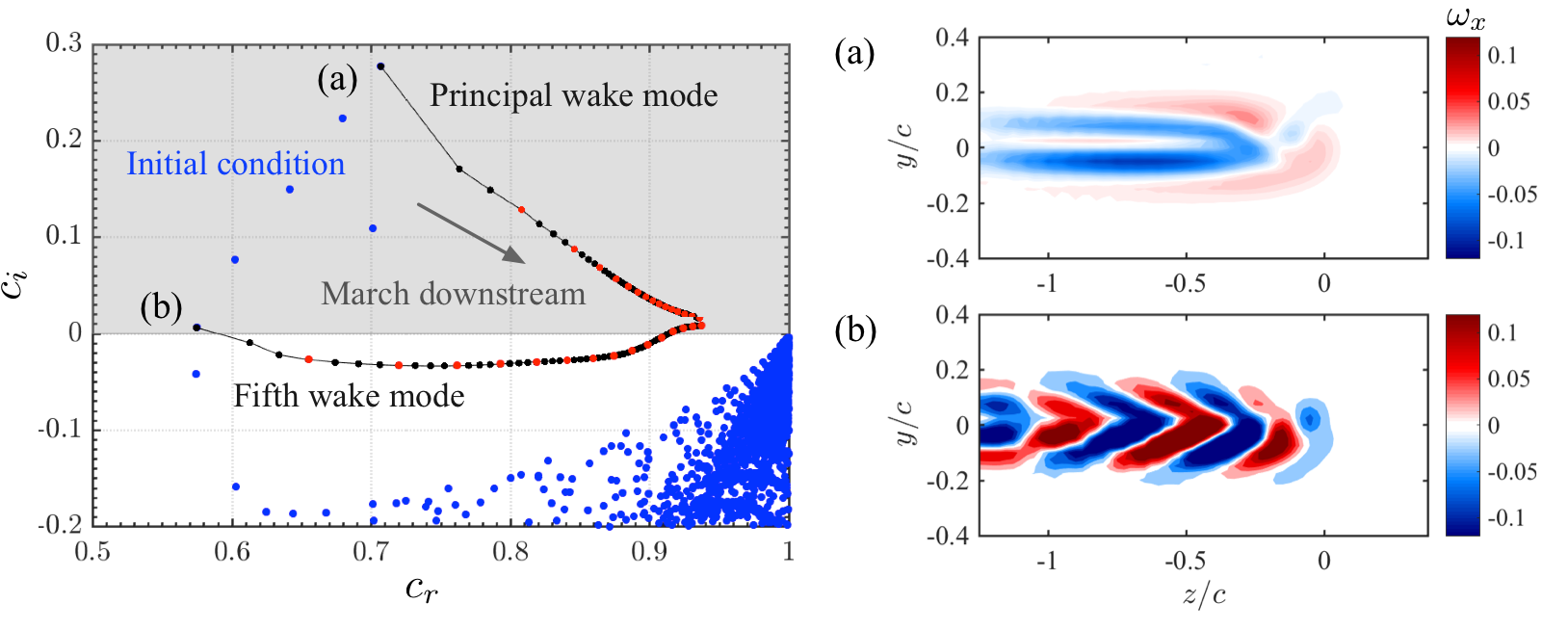}
    \caption{The blue dots represent the eigenvalue spectrum of the
      initial condition at $x/c = 0.25$ with the grey shaded area
      denoting the unstable half plane. The lines represent the change
      in growth rate with downstream progression, with red dots
      indicating the location of integer chord numbers (i.e., $x/c =
      1, 2, ...$). Insets in the figures are the streamwise-vorticity shape functions of
      the (a) principal and (b) fifth wake modes.}
    \label{fig:PSE_eig}
  \end{center}
\end{figure}

The streamwise parabolic nature of the PSE requires an initial condition upstream. This initial condition
is taken from a global stability analysis using a locally-parallel-flow
assumption~\citep[see][]{Edstrand:JFM18}. The parallel
analysis is performed at a streamwise location of $x/c=0.25$ where the
wake deficit is significantly larger than the nascent
vortex. We performed a parametric study varying the frequency to determine the most unstable parallel flow case at $x/c = 0.25$, corresponding to $\omega = 4.5$. Consequently, a wake branch occurs with five unstable
wake-dominated modes as well as a vortex branch containing a single vortex mode as shown by the blue
dots in figure~\ref{fig:PSE_eig}. For further insight into the spectrum, the reader is referred to~\cite{Edstrand:JFM18}. In this work, we focus on
the characteristics of the most unstable mode, termed the principal
wake mode (figure~\ref{fig:PSE_eig}$(a)$), and the mode with the smallest growth rate, the fifth wake mode (figure~\ref{fig:PSE_eig}$(b)$).

\begin{figure}
  \begin{center}
    \includegraphics[width=0.99\textwidth]{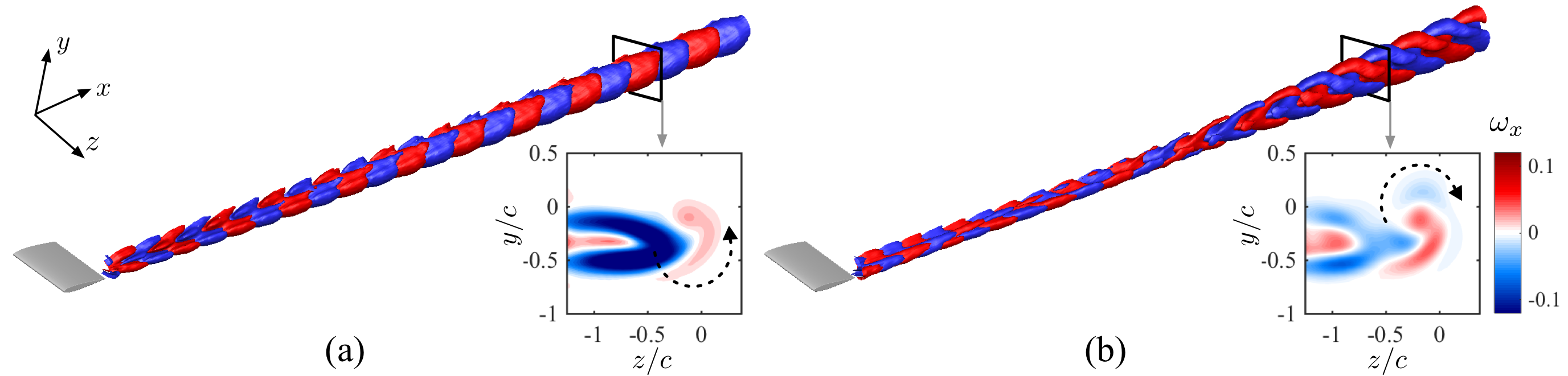}
    \caption{Isosurfaces of streamwise vorticity $\omega_x=\pm 0.008$
      for the (a) principal and (b) fifth wake modes in tip vortex region. The
      insets correspond to a cross-sectional cut of streamwise
      vorticity $\omega_x$. The broken arrows indicate the
      rotation directions of the modes.}
    \label{fig:PSE_modes}
  \end{center}
\end{figure}

In light of our goal to attenuate the trailing vortex, it initially appears
prudent to use the principal (most unstable) mode rather than the mode with the
smallest unstable growth rate. However, observing the progression of the growth rates
as we move downstream (see the black lines in
figure~\ref{fig:PSE_eig}), the growth rate of the principal wake mode
decays monotonically whereas the fifth wake mode becomes increasingly
unstable farther downstream as the base-flow vortex develops. Insight into this eventual instability can be gained by 
observing the progression of each mode shape function. The key observation relates 
to the direction of rotation of
the two modes: the principal mode imposes a counter-clockwise rotation
on the trailing vortex, whereas the fifth mode shows a
clockwise-rotating structure at the same location (see
figure~\ref{fig:PSE_modes} and insets). The eventual 
clockwise nature of the fifth wake mode far downstream is indicative 
of a vortex instability~\citep{Khorrami:1991, Mayer:1992}, whereas 
the counter-clockwise rotation of the principal wake mode is the result 
of base flow field convection. The structure of the vortex instability has 
important implications with respect to flow control. Specifically, 
despite the fact that the principal wake mode has a larger growth rate 
upstream, the fifth wake mode may provide a pathway to excite the vortex
instability farther downstream that may result in the effective attenuation of
the tip vortex.

As such, we hypothesize that excitation of the fifth wake mode will accelerate 
the decay of the trailing vortex. To test this theory, we perform direct 
numerical simulations on the trailing vortex flow field with forcing in the shape of the principal wake mode and the fifth wake mode 
to investigate the effect of the force on the vortex development downstream relative to the baseline 
(no control) case. Both modes are triggered by
localized body forces in the wake near the trailing edge of the
airfoil; a superposition of them enables access to and provides
controllability of the trailing vortex flow. These two modes will thus form
the basis of a control strategy for the breakup of the trailing vortex
flow.

\section{PSE-inspired flow control}

The spatial profile of the fifth mode presents a counter-rotating feature near the tip-vortex core as it evolves downstream.  As such, this mode holds a potential to effectively attenuate the tip vortex. Although this mode is not the most unstable mode, it is known from past free vortex studies that counter-rotating instabilities can efficiently destablize the vortex core~\citep{Khorrami:1991, Mayer:1992}. In this section, we consider using the spatial profiles of the principal and fifth modes as a local forcing input behind the trailing edge, and assess their control effectiveness in reducing the tip vortex strength.   

\begin{figure}
\begin{center}
\includegraphics[width=0.98\textwidth]{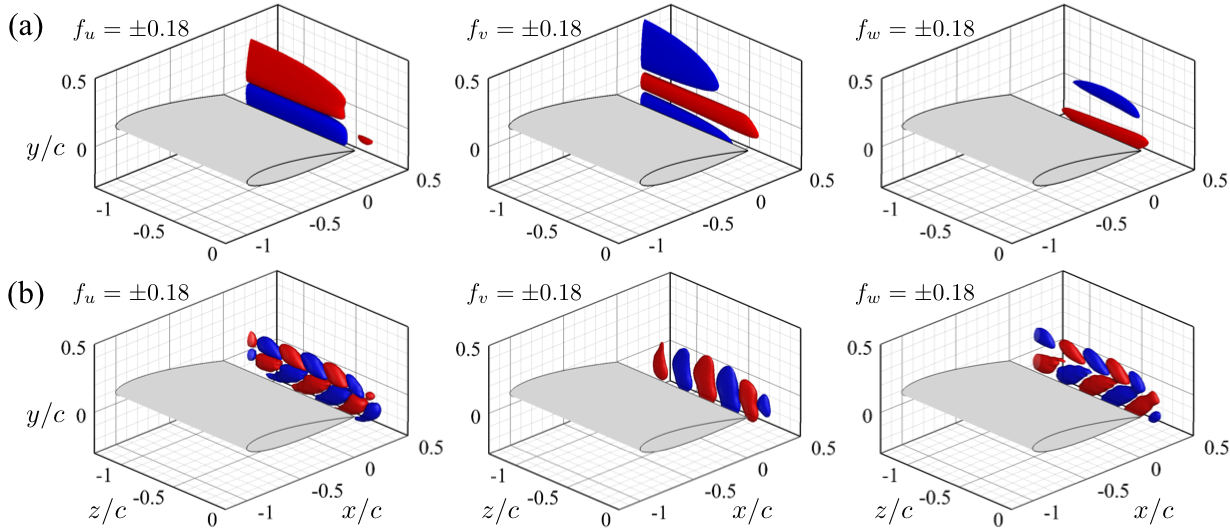}
\caption{Isosurfaces of the 3D body force $f_u$, $f_v$, and $f_w=\pm0.18$ for $x$-, $y$-, and $z$-momentum equations, respectively. The values of isosurfaces are indicated in the figure with a reference of $||\boldsymbol{q}_r||=1$. Blue and red represent negative and positive values, respectively. Body force profiles are based on the (a) principal and (b) fifth wake modes.}
\label{fig:BodyForce}
\end{center}
\end{figure}

The spatial structures of the real part of the shape function from the stability analysis $\boldsymbol{q}_r(\boldsymbol{x}) =\text{Re}\{ \boldsymbol{q}(\boldsymbol{x})\}$ of the principal and fifth modes are substituted into Eq.~(\ref{Eq:BodyForce}) to form the body force actuators, which are visualized in figure \ref{fig:BodyForce}. After implementing the body force in the DNS, we assessed the influence of the forcing amplitude $A$ since it is desirable to keep the forcing input small, while achieving significant attenuation of the tip vortex strength. The value of $A=0.5$ has been verified to result in notable modifications of the base flow. For smaller forcing amplitudes, the added perturbation dissipates without significant attenuation of the tip vortex. 

To quantify the control input, we estimate the oscillatory momentum coefficient $C_\mu$
\begin{equation}
        C_\mu = \frac{\frac{\rho S_\text{act}}{V_\text{act}} \int_{V_\text{act}} 
        [G(x)(\tilde u'^2+\tilde v'^2 + \tilde w'^2)] 
        {\rm d}V}{\frac{1}{2}\rho U_\infty^2 (bc/2)}, 
        \label{Eq:Cmu}
\end{equation}
where $\tilde u'$, $\tilde v'$ and $\tilde w'$ are the fluctuating velocity components, and $S_\text{act}$ and $V_\text{act}$ correspond to where the localized body forcing is active.  Both controlled cases based on the principal and fifth modes yield $C_\mu=0.05$.

\begin{figure}
\begin{center}
\includegraphics[width=0.95\textwidth]{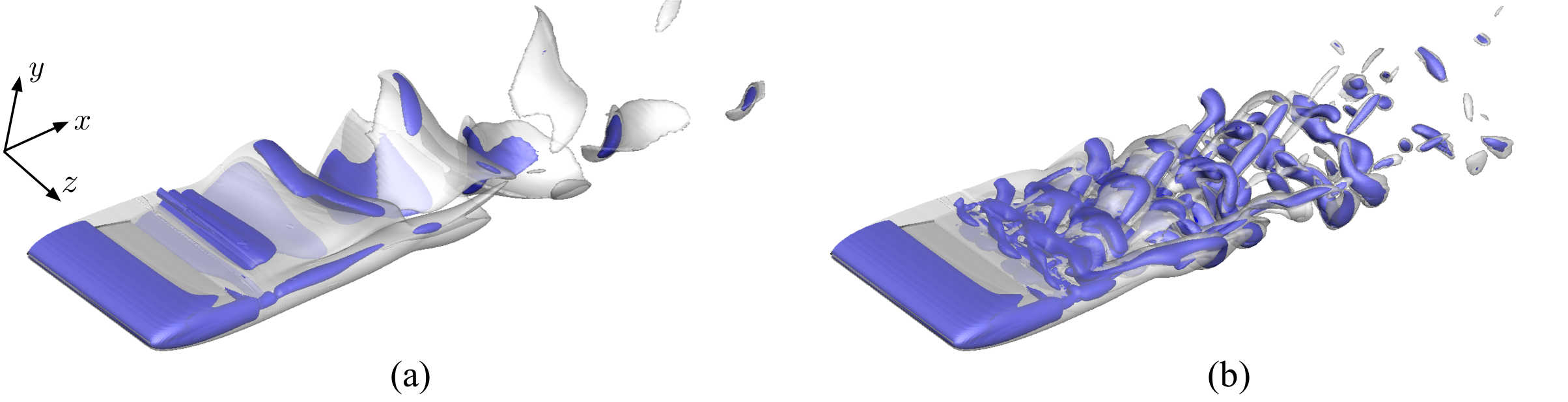}
\caption{Isosurfaces of $Q = 1$ (blue) and vorticity magnitude $\|\boldsymbol{\tilde \omega}\| = 2$ (transparent gray) from instantaneous controlled flow fields using the (a) principal and (b) fifth wake modes as body force.}
\label{fig:Instant}
\end{center}
\end{figure}

The instantaneous flow fields with control applied are shown in figure~\ref{fig:Instant}. The principal mode excites the wake instability in the flow. We observe large fluctuations and vortex shedding downstream of the trailing edge. Due to the unsteadiness introduced by the body force, the tip-vortex core exhibits a moderate level of oscillations. 
In the controlled case based on the fifth mode, the length scale of the body force is finer compared to the controlled case based on the principal mode, leading to the fifth mode case  exhibiting smaller-scale structures. The vertical extent of the wake structure is narrower than the wake of the controlled flow using the principal mode. Furthermore, the tip vortex highly deforms as it convects downstream.

\begin{figure}
\begin{center}
\includegraphics[width=0.97\textwidth]{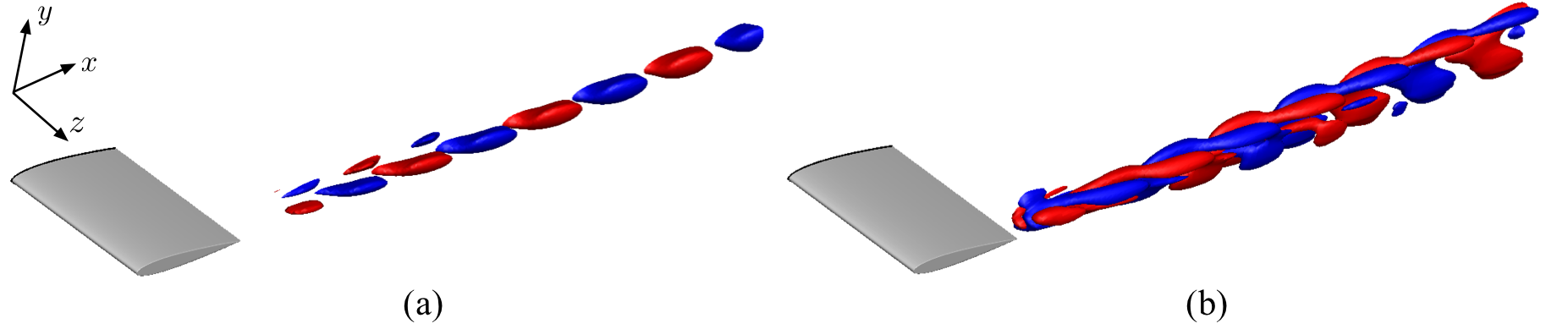}
\caption{Isosurfaces of streamwise-vorticity ($\omega_x'=\pm0.01$) of velocity DMD modes associated with frequency $\omega=4.5$ for controlled cases with the (a) principal and (b) fifth wake modes as body force. The modal structures are visualized only in tip vortex region to highlight the rotation behavior.}
\label{fig:DMD}
\end{center}
\end{figure}

To further analyze the effects of the present control approach, dynamic mode decomposition (DMD) \citep{Schmid:JFM10} is used to extract prominent modes from the controlled flows. We collect two periods of instantaneous velocity field data after the control effects have reached an asymptotic state. For both of the controlled cases, the dominant fluctuation is associated with the control frequency of $\omega=4.5$. For the tip vortex region, the isosurfaces of streamwise vorticity of the dominant mode from each controlled case are presented in figure~\ref{fig:DMD}. 
For the controlled flow with the principal mode, we see perturbations aligned along the streamwise direction. 
For the fifth-wake-mode controlled case, the vorticity perturbations are braided as they convect downstream, and the braiding is in the counter-rotating direction with respect to the base flow, which agrees with the PSE prediction. While the PSE analysis suggests that the counter-rotating feature of the fifth mode becomes prominent past $x/c\approx8$, the nonlinearity in the flow accelerates the emergence of such a counter-rotating feature.  
As the control effort based on the principal wake mode cannot generate the counter-rotating perturbation, the perturbation in the form of the fifth wake mode is indeed required to instigate the correct counter-rotating perturbation around the tip vortex.

\begin{figure}
\begin{center}
\includegraphics[width=1.0\textwidth]{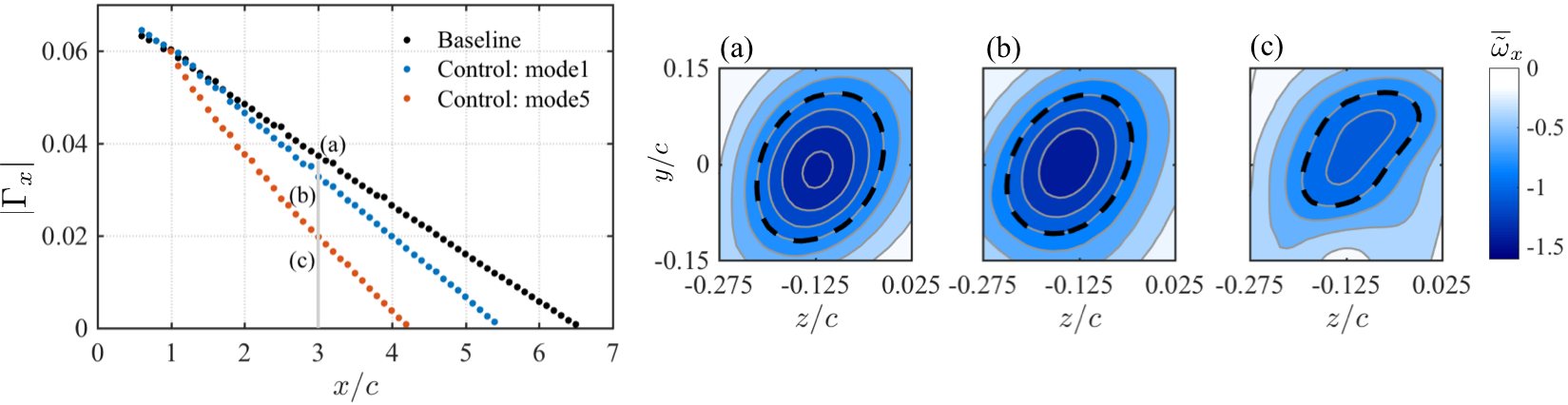}
\caption{The tip vortex circulation $\Gamma_x$ shown for the baseline and controlled cases evaluated at streamwise slices ($y$-$z$ plane). The integration path is indicated by the dashed lines (-~- for $\overline{\tilde{\omega}}_x=-0.8$) in subplots (a) baseline, (b) control using the principal wake mode, and (c) control using the fifth wake mode on the $y$-$z$ plane at $x/c=3$.}
\label{fig:Integration}
\end{center}
\end{figure}

The tip vortex strength is quantified by evaluating the streamwise circulation $\Gamma_x$ using time-averaged velocity distributions $\overline{\boldsymbol{\tilde u}}$ from the baseline and controlled flows. 
From the 3D flow field, we extract 2D slices ($y$-$z$ planes) at various streamwise locations for this analysis. 
For the integration path, a contour line of the time-averaged streamwise vorticity (constant $\overline{\tilde{\omega}}_x$) on each slice is chosen to solely capture the tip vortex structure in both baseline and controlled cases. The tip vortex strength evaluated with a contour of  $\overline{\tilde{\omega}}_x=-0.8$ is shown in figure~\ref{fig:Integration}. 
A comparison of the circulation $|\Gamma_x|$ over $x$ for the baseline and controlled cases suggests that both controlled cases achieve a significant reduction in vortex strength. Noteworthy here is that a larger decrease is realized in the controlled case using the fifth wake mode as the body force.
Although we only report the circulation with the integration path along $\overline{\tilde{\omega}}_x=-0.8$, we have observed that the behavior reported in figure~\ref{fig:Integration} is insensitive to the integration path.

\begin{figure}
\begin{center}
\includegraphics[width=0.7\textwidth]{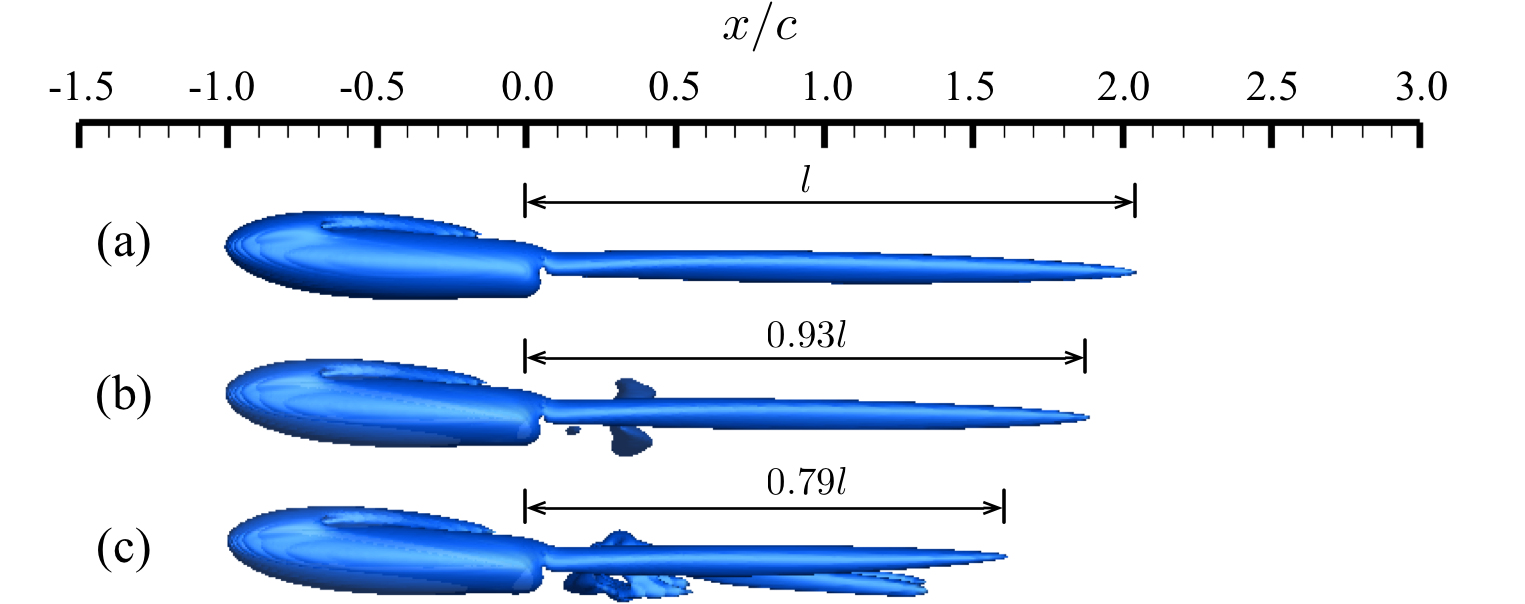}
\caption{Visualization of the tip vortices with iso-surface of $Q=0.8$ (time averaged). Shown are (a) baseline flow, (b) controlled flow using the principal mode, and (c) controlled flow using the fifth mode.}
\label{fig:CoreLength}
\end{center}
\end{figure}

For a more qualitative comparison, we assess the length of the tip vortex core that is visualized using the $Q$-criterion \citep{Hunt:88}, as shown in figure~\ref{fig:CoreLength}. Compared to the baseline vortex core length $l$, the cores for the controlled cases based on the principal and fifth wake modes are shortened by $7\%$ and $21\%$, respectively (based on $Q = 0.8$). There are some secondary structures that also emerge around the tip vortex induced by the actuation force, as shown in figure~\ref{fig:CoreLength}. While the streamwise length of the vortex core varies slightly according to the selected isocontour value of $Q$, the tip vortex length is always observed to diminish for both controlled cases when compared to the baseline flow, regardless of the value of $Q$ used for this assessment.  

While we observe noticeable chances in the wake caused by the control efforts, the time-averaged lift and drag forces for both controlled cases are within 3.2\% of the baseline flow, as shown in table~\ref{table:forces}. The present control schemes are able to deform the tip vortex by leveraging the underlying instability mechanisms. As the deformed tip vortex possesses higher curvature, viscous diffusion and vortex annihilation accelerate the tip vortex attenuation without significantly affecting the forces on the wing. Thus, the present control approach becomes attractive for implementation, as it does not interfere significantly with the aerodynamic performance of the wing itself. Moreover, the fluctuations of the aerodynamic forces in the controlled case based on the principal mode are nearly ten times larger than the fifth-mode-based control case. 

\begin{table}
    \centering
    \begin{tabular}{m{1.65in}m{0.9in}m{0.7in}m{0.9in}m{0.7in}} 
        Cases                       & $C_L$               & $\tilde \sigma_{C_L}$ & $C_D$                 & $\tilde \sigma_{C_D}$  \\ \hline
        Baseline                    & $0.158$             & -   & $0.125$               &   -       \\
        Controlled (principal mode) & $0.160$ ($+1.5\%$)  & 0.00391 & $0.126$ ($+0.8\%$)    &  0.00040       \\
        Controlled (fifth mode)     & $0.163$ ($+3.2\%$)  & 0.00042 & $0.127$ ($+1.6\%$)    &  0.00003      \\
    \end{tabular}
    \caption{Assessment of the control influence on time-averaged lift ($C_L$), drag ($C_D$) coefficients and their standard deviations ($\tilde \sigma$).}
    \label{table:forces}
\end{table}

\section{Conclusion}

The present study utilized a linear parabolized stability analysis as a tool to guide the design of a flow control scheme to attenuate a trailing vortex. Contrary to intuition, the stability analysis of this complex three-dimensional flow field revealed that the vortex instability was excited through the so-called ``fifth wake mode" with an inferior growth rate compared to the most unstable ``principal wake mode." From this, we investigated controlled cases using the principal and fifth wake modes, finding that the fifth wake mode indeed reduces the tip vortex length by $7\%$ and $21\%$, respectively, relative to the uncontrolled reference case. This result provides the insight that, even though the principal wake mode is more unstable, the higher-order wake mode provides a viable pathway to excite the vortex instability and thus yielding a more effective attenuation of the vortex.

From this result, we are able to gain a better understanding of the underlying physical mechanism behind the accelerated attenuation. In particular, we postulate that the fifth wake mode generates counter-rotating streaks, analogous to boundary layer streaks~\citep{Andersson:1998}, that trigger perturbations that counter-rotate relative to the base flow vortex, thereby effectively exciting the instability. Furthermore, we conjecture that disturbing the flow field in the tip-vortex region may be less effective due to the relatively weak nascent vortex development. The complex interaction of the wake and vortex motion provides a quintessential and typical example for the effectiveness of stability analysis (even in approximate form) in guiding the control design for complex flow fields.

\section*{Acknowledgements}

We gratefully acknowledge the support from the US Office of Naval Research (grant number: N00014-15-1-2403). YS and KT thank the Research Computing Center at the Florida State University.

\bibliography{myrefs.bib}

\begin{thebibliography}{21}
\expandafter\ifx\csname natexlab\endcsname\relax\def\natexlab#1{#1}\fi
\def\au#1{#1} \def\ed#1{#1} \def\yr#1{#1}\def\at#1{#1}\def\jt#1{\textit{#1}}
  \def\bt#1{#1}\def\bvol#1{\textbf{#1}} \def\vol#1{#1} \def\pg#1{#1}
  \def\publ#1{#1}\def\arxiv#1{#1}\def\org#1{#1}\def\st#1{\textit{#1}}

\bibitem[Andersson {\em et~al.\/}(1998)Andersson, Henningson \&
  Hanifi]{Andersson:1998}
{\sc \au{Andersson, P}, \au{Henningson, Dan} \& \au{Hanifi, A}} \yr{1998}
  \at{{On a stabilization procedure for the parabolic stability equations}}.
  \jt{Journal of Engineering Mathematics}  \bvol{33},  \pg{311--332}.

\bibitem[Crouch(2005)]{Crouch:2005ce}
{\sc \au{Crouch, Jeffrey}} \yr{2005}  \at{{Airplane trailing vortices and their
  control}}.  \jt{Comptes Rendus Physique}  \bvol{6}~(4-5),  \pg{487--499}.

\bibitem[Devenport {\em et~al.\/}(1996)Devenport, Rife, Liapis \&
  Follin]{Devenport:1996}
{\sc \au{Devenport, William~J}, \au{Rife, Michael~C}, \au{Liapis, Stergios~I}
  \& \au{Follin, Gordon~J}} \yr{1996}  \at{{The Structure and Development of a
  Wing-Tip Vortex}}.  \jt{Journal of Fluid Mechanics}  \bvol{312},
  \pg{67--106}.

\bibitem[Edstrand \& Cattafesta(2015)]{Edstrand:2015}
{\sc \au{Edstrand, Adam} \& \au{Cattafesta, Louis~N}} \yr{2015} {Topology of a
  Trailing Vortex Flow Field with Steady Circulation Control Blowing}.  \bt{In
  {\em 53rd AIAA Aerospace Sciences Meeting\/}},  \pg{pp. 1--15}.
  \publ{Kissimmee, Florida: American Institute of Aeronautics and
  Astronautics}.

\bibitem[Edstrand {\em et~al.\/}(2016)Edstrand, Davis, Schmid, Taira \&
  Cattafesta]{Edstrand:2016}
{\sc \au{Edstrand, Adam~M}, \au{Davis, Timothy~B}, \au{Schmid, Peter~J},
  \au{Taira, Kunihiko} \& \au{Cattafesta, Louis~N}} \yr{2016}  \at{{On the
  mechanism of trailing vortex wandering}}.  \jt{Journal of Fluid Mechanics}
  \bvol{801},  \pg{R1--11}.

\bibitem[Edstrand {\em et~al.\/}(2018)Edstrand, Schmid, Taira \&
  Cattafesta]{Edstrand:JFM18}
{\sc \au{Edstrand, A.~M.}, \au{Schmid, P.~J.}, \au{Taira, K.} \&
  \au{Cattafesta, L.~N.}} \yr{2018}  \at{A parallel stability analysis of a
  trailing vortex wake}.  \jt{J. Fluid Mech.}  \bvol{837},  \pg{8585--895}.

\bibitem[Ham \& Iaccarino(2004)]{Ham:CTR04}
{\sc \au{Ham, F.} \& \au{Iaccarino, G.}} \yr{2004}  \bt{Energy conservation in
  collocated discretization schemes on unstructured meshes}.  \pg{pp. 3--14}.
  {Annual Research Brief, Center for Turbulence Research, Stanford Univ.}

\bibitem[Ham {\em et~al.\/}(2006)Ham, Mattsson \& Iaccarino]{Ham:CTR06}
{\sc \au{Ham, F.}, \au{Mattsson, K.} \& \au{Iaccarino, G.}} \yr{2006}
  \bt{Accurate and stable finite volume operators for unstructured flow
  solvers}.  \pg{pp. 243--261}. {Annual Research Brief, Center for Turbulence
  Research, Stanford Univ.}

\bibitem[Herbert(1997)]{Herbert:1997}
{\sc \au{Herbert, Thorwald}} \yr{1997}  \at{{Parabolized Stability Equations}}.
   \jt{Annual Review of Fluid Mechanics}  \bvol{29},  \pg{245--283}.

\bibitem[Hunt {\em et~al.\/}(1988)Hunt, Wray \& Moin]{Hunt:88}
{\sc \au{Hunt, J. C.~R.}, \au{Wray, A.~A.} \& \au{Moin, P.}} \yr{1988}
  \at{Eddies, streams, and convergence zones in turbulent flows}.  \jt{Proc. of
  the Summer Program, Center of Turbulence Research}  \pg{pp. 193--208}.

\bibitem[Jacquin {\em et~al.\/}(2001)Jacquin, Fabre, Geffroy \&
  Coustols]{Jacquin:2001}
{\sc \au{Jacquin, Laurent}, \au{Fabre, David}, \au{Geffroy, P} \& \au{Coustols,
  E}} \yr{2001} {The Properties of a Transport Aircraft Wake in the Extended
  Near Field: an Experimental Study}.  \bt{In {\em th Aerospace Sciences
  Meeting\/}},  \pg{pp. 1--42}.

\bibitem[Khorrami(1991)]{Khorrami:1991}
{\sc \au{Khorrami, Mehdi~R}} \yr{1991}  \at{{On the viscous modes of
  instability of a trailing line vortex}}.  \jt{Journal of Fluid Mechanics}
  \bvol{225},  \pg{197--212}.

\bibitem[Margaris \& Gursul(2006)]{Margaris:2006}
{\sc \au{Margaris, P} \& \au{Gursul, I}} \yr{2006}  \at{{WIng tip vortex
  control using synthetic jets}}.  \jt{The Aeronautical Journal}
  \bvol{110}~(1112),  \pg{673--681}.

\bibitem[Margaris \& Gursul(2010)]{Margaris:2010}
{\sc \au{Margaris, P} \& \au{Gursul, I}} \yr{2010}  \at{{Vortex topology of
  wing tip blowing}}.  \jt{Aerospace Science and Technology}  \bvol{14},
  \pg{143--160}.

\bibitem[Matalanis \& Eaton(2007)]{Matalanis:2007}
{\sc \au{Matalanis, Claude~G} \& \au{Eaton, John~K}} \yr{2007}  \at{{Wake
  Vortex Alleviation Using Rapidly Actuated Segmented Gurney Flaps}}.  \jt{AIAA
  Journal}  \bvol{45}~(8),  \pg{1874--1884}.

\bibitem[Mayer \& Powell(1992)]{Mayer:1992}
{\sc \au{Mayer, Ernst} \& \au{Powell, Kenneth}} \yr{1992}  \at{{Viscous and
  inviscid instabilities of a trailing vortex}}.  \jt{Journal of Fluid
  Mechanics}  \bvol{245},  \pg{91--114}.

\bibitem[Schmid \& Henningson(2001)]{Schmid:2001}
{\sc \au{Schmid, Peter} \& \au{Henningson, Dan}} \yr{2001} {\em {Stability and
  Transition in Shear Flows}\/}.  \publ{Springer}.

\bibitem[Schmid(2010)]{Schmid:JFM10}
{\sc \au{Schmid, Peter~J.}} \yr{2010}  \at{Dynamic mode decomposition of
  numerical and experimental data}.  \jt{J. Fluid Mech.}  \bvol{656}~(10),
  \pg{5--28}.

\bibitem[Spalart(1998)]{Spalart:1998dl}
{\sc \au{Spalart, Philippe~R}} \yr{1998}  \at{Airplane trailing vortices}.
  \jt{Annual Review of Fluid Mechanics}  \bvol{30}~(1),  \pg{107--138}.

\bibitem[Streett {\em et~al.\/}(2003)Streett, Lockard, Singer, Khorrami \&
  Choudhari]{Streett:2003vx}
{\sc \au{Streett, C~L}, \au{Lockard, D~P}, \au{Singer, B~A}, \au{Khorrami, M~R}
  \& \au{Choudhari, M~M}} \yr{2003} {\em {In Search of the Physics: The
  Interplay of Experiment and Computation in Airframe Noise Research: Flap-Edge
  Noise}\/}.

\bibitem[Taira \& Colonius(2009)]{Taira:AIAAJ09}
{\sc \au{Taira, K.} \& \au{Colonius, T.}} \yr{2009}  \at{Effect of tip vortices
  and low-{R}eynolds-number poststall flow control}.  \jt{AIAA Journal}
  \bvol{47}~(3),  \pg{749--756}.

\end{thebibliography}
\bibliographystyle{jfm}

\end{document}